\documentclass[9pt,twoside]{rmaa-rho-class/rmaa-rho}

\RMxAAtemplatetype{\RMxAA}

\vol{61}              
\pages{1-10}          
\thisyear{2025}       
\doi{https://www.astroscu.unam.mx/rmaa/RMxAA..ELLIPSECT} 

\usepackage[utf8]{inputenc}
\usepackage{listings}
\usepackage{xcolor}

\DeclareMathOperator{\sech}{sech}

\title{\texttt{EllipSect}: A Surface Brightness Analysis Tool for \texttt{GALFIT 3}}

\author[1]{C.~A\~norve \orcidlink{0000-0002-3721-8869}}
\author[2]{O.~U.~Reyes-Amador \orcidlink{0000-0001-7707-7389} }
\author[3,4,5]{E.~R\'ios-L\'opez  \orcidlink{0000-0002-4436-221X}}
\author[6]{D.~de~Ram\'on~Tadeo}
\author[3]{O.~L\'opez-Cruz  \orcidlink{0000-0002-1381-7437}}

\affil[1]{Facultad de Ciencias de la Tierra y el Espacio, Universidad Aut\'onoma  de Sinaloa, Blvd. de la Am\'ericas y Av. Universidad s/n, Ciudad Universitaria,  Culiac\'an, Sinaloa, 80013, M\'exico.}
\affil[2]{Instituto de Radioastronom\'ia y Astrof\'isica, UNAM, Campus  Morelia, AP 3-72, C.P. 58089, M\'exico.}
\affil[3]{Instituto Nacional de Astrof\'isica, \'Optica y Electr\'onica  (INAOE), Luis Enrique Erro 1, Apartado Postal 51 y 216,  C.P. 72840 Puebla, M\'exico.}
\affil[4]{Instituto de Astrof\'isica de Canarias (IAC), V\'ia L\'actea s/n, 38205, La Laguna, Tenerife, Espa\~na.}
\affil[5]{Universidad Pedag\'ogica del Estado de Sinaloa, Castiza s/n, 80027, Culiac\'an, Sinaloa, M\'exico.}
\affil[6]{Facultad de Ciencias de la Computaci\'on, Benem\'erita Universidad Aut\'onoma de Puebla (FCC-BUAP), Av. San Claudio y 14 Sur, Ciudad Universitaria, C.P. 72570, Puebla, M\'exico.}

\leadauthor{A\~norve et al.}
\smalltitle{\texttt{EllipSect}: A Surface Brightness Tool for GALFIT 3}

\corres{Christopher A\~norve}
\email{canorve@uas.edu.mx}

\keywords{galaxies: photometry, galaxies: fundamental parameters, methods: data analysis }

\setbool{rho-abstract}{true} 
\setbool{rho-resumen}{true} 

\begin{abstract}
\texttt{EllipSect} is a user-friendly analysis and measurement tool, implemented in \texttt{Python},
that operates on the imaging data together with the output of the widely used 2D surface-brightness
fitting code \texttt{GALFIT 3}. It produces publication-quality figures and exportable data products
to enable quantitative assessment of \texttt{GALFIT 3} models and their individual components. In addition,
\texttt{EllipSect} computes non-parametric measurements that are not provided by \texttt{GALFIT 3}, including
the total effective radius resulting from multi-component fits, cusp radius, and the Petrosian radius.
This paper provides examples and a quick guide for \texttt{EllipSect}.
\end{abstract}

\begin{resumen}
\texttt{EllipSect} es una herramienta de an\'alisis y medici\'on, implementada en \texttt{Python},
que opera sobre los datos de imagen junto con la salida del c\'odigo ampliamente utilizado de
ajuste bidimensional de brillo superficial \texttt{GALFIT 3}. Genera figuras de calidad publicable
y productos de datos exportables para permitir una evaluaci\'on cuantitativa de los modelos de
\texttt{GALFIT 3} y de sus componentes individuales. Adem\'as, \texttt{EllipSect} calcula mediciones
no param\'etricas que no son proporcionadas por \texttt{GALFIT 3}, incluyendo el radio efectivo total
resultante de ajustes multicomponente, el radio de c\'uspide y el radio de Petrosian. Este art\'iculo
presenta ejemplos y una gu\'ia r\'apida de uso de \texttt{EllipSect}.
\end{resumen}

\begin{document}

\nolinenumbers

\maketitle
\pagestyle{fancy}\thispagestyle{firststyle}

\section{Introduction}
\label{sec:intro}

 \texttt{GALFIT 3}\footnote{For installation, usage, and general help, visit 
 \url{https://users.obs.carnegiescience.edu/peng/work/galfit/galfit.html}} 
 \citep[][]{peng02,peng10} is a widely used stand-alone two-dimensional (2D) fitting code designed to model the surface brightness (SB) distribution of astronomical sources by generating 2D-parametric models that account for galaxies' substructures and individual stars, among others. With \texttt{GALFIT 3}, the user can apply multiple profiles to fit the following components: bulges, disks, bars, spiral arms, and asymmetries to one or several galaxies simultaneously.
 
 Furthermore, when dealing with multiple components, the orientation of each component is independent of the others; hence, component-based separation is more reliable.  This property gives a substantial advantage in crowded galaxy fields, such as  compact groups or clusters of galaxies. Indeed, our experience shows that masking does not adequately account for light from nearby or overlapping galaxies; instead, fitting the galaxy of interest along with its neighboring galaxies yields a better result. Indeed,  \citet{haussler07, anorve12} tested this recommendation using simulated data and found that fitting neighboring galaxies simultaneously accounts for galaxies' extended light more precisely.  

 Another feature of \texttt{GALFIT 3} is that a user-supplied point spread function (PSF) image is necessary, or the user may utilize the built-in functions provided by \texttt{GALFIT 3} to model the PSF. \texttt{GALFIT 3} convolves the PSF with the chosen profile to fit a given galaxy. This differs from some previous methods, which relied on deconvolving the image \citep[for example][]{lucy74, richardson72} before generating isophotes \citep[e.g.,][]{jedrzejewski87} and fitting the SB profile.

\texttt{GALFIT 3} applies a $\chi^2_\nu$ optimization scheme for parameter selection, 
including the background, generating parameter uncertainties, and producing a FITS 
(Flexible Image Transport System) image cube containing the image of the input object, 
the model image resulting after the fit, and the residual image. However, \texttt{GALFIT 3} 
does not include a graphical tool to compare the input image with its resulting model 
image. Users commonly applied one-dimensional (1D) IRAF's \texttt{ellipse} and other 
plotting programs; nevertheless, this suite of programs became difficult to access 
after IRAF's discontinuation in 2013. Fortunately, IRAF's \texttt{ellipse} is 
implemented in the Astropy library and is available in new releases. In any case, 
the main difficulty is that one must convert \texttt{GALFIT 3}'s output 
format to a format accepted by external codes. Nevertheless, this process could be 
time-consuming and error-prone. However, this method is not feasible with IRAF's 
\texttt{ellipse} task, which processes only one galaxy at a time. Also, a common misconception suggests that 1D approaches work better in low signal-to-noise (S/N) regions or the presence of low-SB features; however, there is no advantage from 1D over the 2D approaches due to the Poissonian nature of the signal\footnote{See a more detailed discussion at 
\url{https://users.obs.carnegiescience.edu/peng/work/galfit/TFAQ.html\#misconception1} on 1D vs. 2D modeling} \citep[see also][]{press2007,peng10}.

 Hence, we are introducing \texttt{EllipSect}, a new algorithm that 
 generates SB profiles and extracts complementary information directly from 
 \texttt{GALFIT 3} 's output, with the current implementation coded 
 in \texttt{Python}. The current manuscript focuses on \texttt{EllipSect} as a post-fit analysis tool
that produces quantitative measurements, publication-quality graphics, and exportable data products
from the observed galaxy and \texttt{GALFIT 3} models. These products can also support iterative model
refinement (e.g., adding or removing components) when needed.
 \texttt{EllipSect} 's output includes data 
 from the observed galaxy, and its model fits in the form of one-dimensional 
 profiles. For a more straightforward assessment of the results, 
 \texttt{EllipSect} can handle individual model components for a detailed analysis.

Additionally, \texttt{EllipSect} provides non-parametric measurements derived from the data
and the \texttt{GALFIT 3} model/output, such as integrated magnitudes, mean 
 SB at $r_e$ ($\left<\mu\right>_e$), 
 the SB at $r_e$ ($\mu_e$), the radius at 90\% of total light, Kron 
 radius \citep{kron80}, Petrosian radius  \citep{petrosian76}, 
 bulge-to-total ratio, and the total effective radius ($r_e^t$) of a galaxy with multiple components fits. 

 To make this paper self-contained, Appendix A provides a brief history, applications, 
and analytical profiles implemented in \texttt{GALFIT 3}. We intended to order them logically, 
from the more general instance through their particular cases, to avoid the impression that 
the suite of profiles included in \texttt{GALFIT 3} is merely a simple list of unrelated profiles.

This paper contains the following sections: \S1 presents a general introduction to \texttt{GALFIT 3} and the built-in profiles; \S2 explains \texttt{EllipSect}'s implemented algorithms. \S3 
shows applications and examples, \S4 and \S5 presents a discussion and conclusions, respectively. The appendices show the build-up of the \texttt{GALFIT 3} functions and explain how to install and run \texttt{EllipSect}.

\section{{\texttt{EllipSect}}}

This section explains how {\texttt{EllipSect}} works. We have divided it into two parts: first, we describe the main algorithm, and then, we explain the photometric measurements. Throughout this article, we refer to the individual profiles that comprise a SB model as the components. In contrast, the set of SB components constitutes the \textit{model}.

\subsection{The Algorithm}
\label{algo}
{\texttt{EllipSect}} was designed to be user-friendly for most users. It omits any direct interaction with the code or translation of \texttt{GALFIT 3} 's data format.  \texttt{EllipSect} is executed from the command line. It requires only the latest \texttt{GALFIT 3} model output files.

\texttt{GALFIT 3}'s output files consist of a data cube in FITS format, which includes the original image (chopped into the fit region), the model image, and the residual image (i.e., galaxy image $-$ model image), as well as a file named {\tt galfit.nn} (where {\tt nn} indicates the number of attempts run by \texttt{GALFIT 3}). This last file contains the parameters that fit each component of the model. Additionally, \texttt{GALFIT 3} creates a log file that includes the model parameter values and their associated errors. \texttt{EllipSect} uses the \texttt{GALFIT} output file ({\tt galfit.nn}) to parse the input magnitude zero point, plate scale, mask file name, output file, and model parameters. 

The output \texttt{GALFIT 3} file may include information about the simultaneous fits of neighboring galaxies to the target galaxy. Using the distances among the model components, \texttt{EllipSect} can identify the elements associated with each galaxy.

\texttt{EllipSect} uses the same mask employed by \texttt{GALFIT} to remove the undesired pixels 
from the image. Once this is done, \texttt{EllipSect} reads the sky component from the \texttt{GALFIT} 
output file and subtracts it from the output galaxy and model images. The program divides the galaxy 
and its model into sectors equally spaced in angle around their centers, using the
\texttt{sectors\_photometry} routine from the \texttt{mgefit}\footnote{\texttt{mgefit} is a software 
to obtain Multi-Gaussian Expansion (MGE) fits of galaxy images \citep{cappellari02}} \texttt{Python} 
library by \citet{cappellari02}. With this routine, \texttt{EllipSect} employs 
these sectors to compute the surface brightness at multiple radii.
It divides an ellipse, centered on the galaxy, into multiple sectors according to the geometric information from the \texttt{GALFIT 3} output file until reaching a minimum count level, set by the \textit{MINLEVEL} parameter, and its preselected value is zero. {\texttt{EllipSect}} takes the angular orientation and ellipticity from the last component of the \texttt{GALFIT 3} output file (see \ref{a2}) as default values. This component is typically an exponential model, although this may vary with the specific galaxy model employed. The user can change all default parameters.

{\texttt{EllipSect}} generates publication-quality graphs. It produces two basic graphs in its simplest mode: one showing the averaged surface 
brightness along the major axis and another displaying the SB for 
various azimuthal angles across multiple plots (see examples in the Results 
section). Additionally, this mode generates a \texttt{png} image featuring 
the galaxy, model, and residual. {\texttt{EllipSect}} can also export a file 
containing the data points for each plot, allowing users to create custom 
visualizations with other plotting routines. Optionally, the SB profile for 
each plot component can include the galaxy (via the \texttt{--comp} flag). This feature allows users to visualize the behavior of model components and identify irrelevant components or redundancies (see, for example, Figure \ref{figa85}).

According to the \texttt{mgefit} manual, the SB results from the following expression\footnote{This does not include the extinction parameter}:

\begin{equation}
  \label{mgeeq}
  \mu_I = zpt + 5\times \log(Pl_{sce}) + 2.5\times \log(t_{exp}) - 2.5\times \log(C_0) + 0.1 ,
\end{equation}

where $zpt$ is the magnitude zero point,  $C_0$ are the $counts\times pixel^{-1}$, $Pl_{sce}$ is the plate scale, $t_{exp}$ is the exposition time, and $0.1$ is a correction for infinite aperture.

\subsection{Output photometry}

In addition to generating surface brightness plots, {\texttt{EllipSect}} calculates additional photometric variables from the models beyond those provided by \texttt{GALFIT 3}. These variables are stored in an output file and are computed after \texttt{EllipSect} extracts the counts for each sector. This section describes these variables and explains how they are calculated.

{\texttt{EllipSect}} connects to NED through the command \texttt{curl} to download an \texttt{xml} file to extract the photometric corrections and compute the absolute magnitude according to the appropriate band. To transform magnitudes to luminosities, and 
vice-versa, the absolute magnitude of the Sun from \citet{willmer18} is assumed. This file extracts the distance modulus and the galactic extinction \citep{schlafly11}. It also retrieves a conversion parameter used to convert angular units to Kpc. SB dimming also corrects the mean SB at effective radius $\left<\mu\right>_e$ and SB at the effective radius $\mu_e$. The only correction not applied is the k-correction, as it needs the color information that is not always available \citep[e.g., see][]{chilingarian12}. It is important to note that corrections are applied only to the quantities in the output file, while the single and multiple plots mentioned earlier 
do not include these corrections. However, users can apply a correction factor either as a command-line argument or by generating plots from the {\texttt{EllipSect}} output files.

NED provides two distance-modulus measurements: one derived from cosmological models and the other obtained directly from the galaxy. By default, 
{\texttt{EllipSect}} uses the redshift-independent measurement to calculate 
the luminosity, if available. Otherwise, it uses the distance modulus based 
on cosmological models.

The majority of photometric parameters are calculated within an ellipse that encloses 90\% of the total light. Some of those associated photometric parameters are the Bumpiness \citep{blakeslee06}, the Tidal parameter \citep{tal09}, the reduced chi-squared ($\chi^2_\nu$) within this ellipse (as opposed to the value provided by \texttt{GALFIT 3}, which applies to the entire fitted image section), the signal-to-noise ratio (S/N), the residual sum of squares, the number of degrees of freedom, the number of free parameters, the absolute magnitude, the bulge-to-total ratio, and the luminosity.

The Bumpiness parameter quantifies the irregular structure of a galaxy by measuring deviations in its SB from a smooth SB model. \citet{blakeslee06} employed this parameter and the S\'ersic index to classify galaxies into elliptical, S0, spiral, and irregular types.

The calculation of the Bumpiness $B$ results from the following formula:

\begin{equation}
\label{bump}
  B = \frac{\left(\frac{1}{N}\sum_{r_{i,j}<2r_e}[I_{i,j}-S(x_i,y_j|r_e,n)]_s^2 - \sigma_s^2\right)^{\frac{1}{2}}}
  {\frac{1}{N}\sum_{r_{i,j}<2r_e}S(x_i,y_j|r_e,n)},
\end{equation}

The sum extends over two effective radii ($2r_e$), but in this case, the sum is taken within the region of the ellipse defined above. Here, $N$ represents the number of unmasked pixels included in the computation, $I_{i,j}$ is the galaxy intensity at pixel $(i,j)$, $S$ is the convolved surface brightness model, and $\sigma_s^2$ is the estimated photometric error, which is taken directly from the sigma image produced by \texttt{GALFIT 3}.

Another useful parameter is the Tidal parameter, defined as:

\begin{equation}
\label{tidal}
  T_{gal} = \left[\left(\frac{I_{(x,y)}}{M_{(x,y)}}\right) - 1\right],
\end{equation}

where $I_{(x,y)}$ are the galaxy pixel intensity values in (x, y)
and $M_{(x,y)}$ represent the intensity values of the model.

This parameter measures the differences between 
the galaxy image and its corresponding model.
\citet{tal09} used this to measure the tidal disturbances in elliptical galaxies and explored their relation with environmental properties, AGN activity, and color.

\texttt{EllipSect} is particularly useful to determine the total effective radius, $r_e^t$, of a galaxy when its SB model results after a multicomponent fit, for example, for a fit including a bulge, a disk, and a bar (S\'ersic+Exponential+S\'ersic), while \texttt{GALFIT 3} only generates $r_e$ for each component; \texttt{EllipSect} calculates the total effective radius by summing all subcomponents of the galaxie's SB model, calculating half of the total flux, and finding the corresponding radius using the bisection method. In the manual of  \texttt{GALFIT 3}, 
the author suggests the application of a single S\'ersic fit for the entire galaxy, using $r_e$ from 
the \texttt{GALFIT 3} fit as a proxy for $r_e^t$. We argue that \texttt{EllipSect} significantly extends this approximation.

To resolve this, \texttt{EllipSect} employs a numerical method that adds together the models of all components and then locates the radius where the total luminosity reaches half of its overall value ($r_e^t$). This procedure can also be applied to determine the radius corresponding to any percentage of the total light.

{\texttt{EllipSect}} provides additional ways to determine the goodness of fit by calculating the Akaike Information Criterion \citep[AIC,][]{akaike74} and the Bayesian Information Criterion \citep[BIC,][]{schwarz78}, which enable to determine the best fit for the galaxy among a given set of models. Both criteria penalize adding new parameters by increasing the estimate of $\chi_\nu$. The model with the lowest AIC or BIC value is considered the best fit. {\texttt{EllipSect}} also computes an alternative BIC using the number of resolution elements instead of the number of pixels defined by $N_{res} = \frac{n_{pix}}{\pi (\text{FWHM})^2}$.

Finally, additional output parameters are computed for each model component, including the fraction of 
light (such as the bulge-to-total ratio for each component), $\left<\mu\right>_e$, $\mu_e$, absolute 
magnitude, luminosity radius at 90\% of the total light, Kron radius \citep{kron80}, and Petrosian 
radius \citep{petrosian76}. Accurate determination of those parameters requires a precise estimation 
of the sky background.

\subsubsection{Sky computation}

An incorrect sky background estimation can lead to inaccurate surface brightness model parameters, especially the S\'ersic index. An appropriate strategy is to determine and fix the sky background during model fitting, thereby reducing the number of free parameters and preventing parameter degeneracy.

\texttt{EllipSect} provides two methods for computing the sky background: the sky gradient method and the random box method. \texttt{GALFIT 3} does not include these two methods; therefore, by comparing the results, the user decides whether to retain the model or refine it by using the newly derived sky value from \texttt{EllipSect}.

\textbf{Sky gradient method}:  This method follows \citet{barden12}. In this approach, {\texttt{EllipSect}} constructs concentric rings using the same angular parameters and ellipticity employed for the elliptical sectors described in \ref{algo}. The user specifies the ring width and {\texttt{EllipSect}} enlarges the rings' major axes outward from the galaxy's center. {\texttt{EllipSect}} can remove the top $80\%$ and bottom $20\%$ of sky values in each ring; however, the user can also specify a mask file to exclude undesired pixels. Finally, {\texttt{EllipSect}} computes the gradient from the mean sky values of each ring.

\begin{figure}[!t]
  \includegraphics[width=\columnwidth]{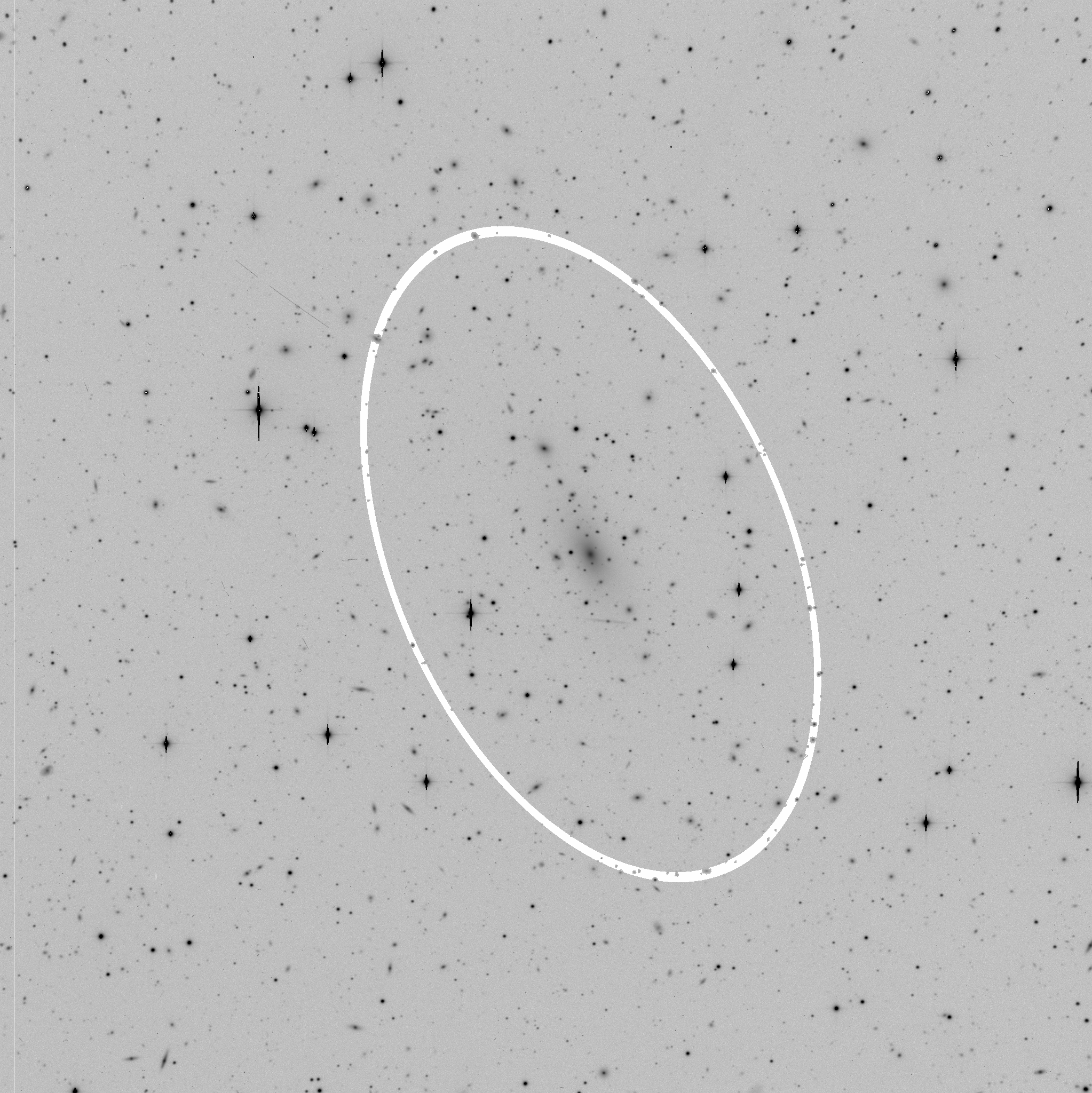}
  \caption{Example of sky computation using the gradient method. The ring around the brightest cluster member of Abell 2029 indicates the radius used by \texttt{EllipSect} to compute the sky. For visualization
purposes, all pixels in that elliptical ring are assigned the same numerical value corresponding to the ellipse's major-axis length.}
\label{fig:simple}
\end{figure}

A negative gradient arises in the galaxy's inner regions because the central counts exceed those in the outskirts. Beyond this zone, the gradient fluctuates randomly between positive and negative values (see Figure 5 in \citet{barden12}). {\texttt{EllipSect}} generates a fit image with sky rings to illustrate this gradient. It begins calculating the gradient from an initial radius and stops when it becomes positive for the second time, following the recommendation by \citet{barden12}. At that point, 
{\texttt{EllipSect}} returns the sky value determined for this radius. Figure \ref{fig:simple} shows an example of an elliptical ring used to compute the sky.

\textbf{Random box method}: For this method, {\texttt{EllipSect}} calculates the sky using randomly placed boxes around the galaxy. To achieve this, it first creates an elliptical mask that covers the entire galaxy. By default, the geometry for this elliptical mask comes from the \texttt{GALFIT 3} output file, although the user may modify it. 

Next, the program positions boxes randomly between this elliptical mask and a specified maximum radius (often at the image boundary). It can also optionally remove the top 80\% and bottom 20\% of pixel values within these boxes. The final sky value is the mean of the median and standard deviation of all pixels within the boxes. The default settings place 20 boxes, each measuring 20 pixels across. This method follows the approach described in \citet{Gao17}.

\section{Results: Examples of SB plots}

\citet{rios21} and \citet{rios25} applied \texttt{EllipSect} in their study of Two Micron All Sky Survey (2MASS) galaxies, where they applied \texttt{GALFIT 3} to generate estimates of morphological and structural parameters through photometric decompositions. In another study by \citet{reyes21}, {\texttt{EllipSect}} was used to map dust in the cores of local Active Galactic Nuclei (AGN), using Hubble Space Telescope (HST) images, SB models, and radiative transfer simulations. The code has also been used in reports, conference proceedings, professional papers, and undergraduate and graduate theses \citep[e.g.,][Valerdi M. et al. 2026, in preparation]{rios21b,duggal24}.

Here, we present examples of SB profiles for three images: Holm 15A from the Kitt Peak National Observatory (KPNO), M61 from 2MASS, and NGC 4261 from the HST.

\subsection{Kitt Peak: Holm~15A}

\begin{figure}[!t]
  \includegraphics[width=\columnwidth]{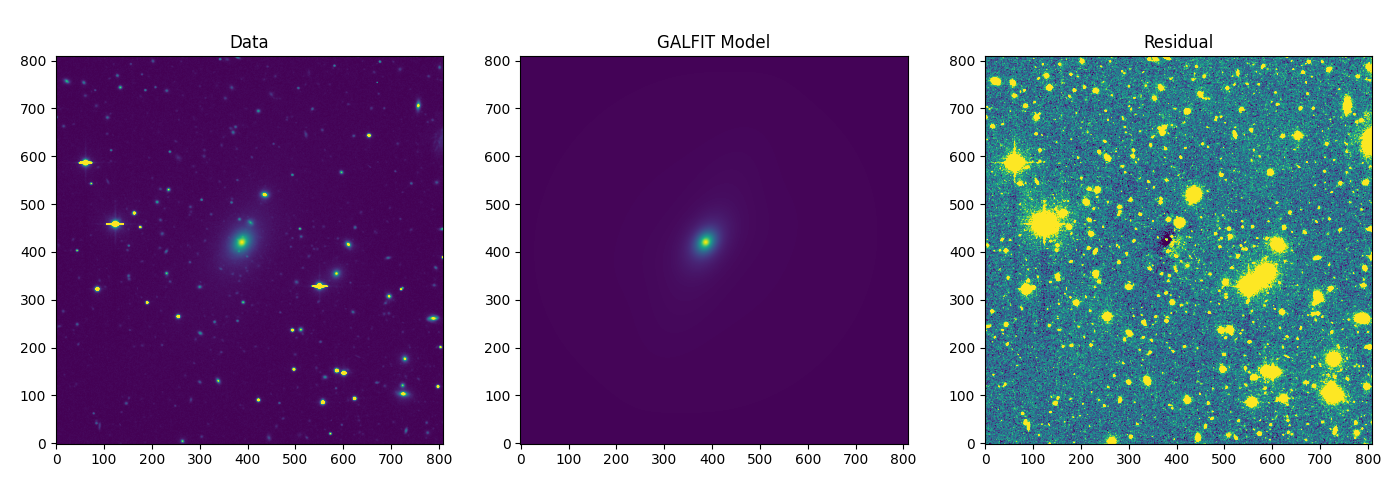}
  \caption{Example of cube image generated by {\texttt{EllipSect}}. The left panel shows the galaxy, the middle panel is the model, and the right panel is the residual.
  The user can change the color palette, brightness, and contrast.}
\label{cube}
\end{figure}

This example shows the SB profile plots of the Abell 85  brightest cluster galaxy (BCG), the cD galaxy Holm~15A. Here, \texttt{GALFIT 3} was used to fit a multi-Gaussian component to model the SB; this result complements the analysis of \citet{2014ApJ...795L..31L}, who used Nuker profiles. 
These observations were obtained with the KPNO 0.9m Telescope and are part of the LOCOS catalog \citep{1997PhDT.......333L}. The reader can find details on the selection criteria, observations, and data reduction in \citet{1997PhDT.......333L}, \citet{2004ApJ...614..679L}, and L\'opez-Cruz et al. (2026, in preparation). In Figure \ref{cube}, we show the cube image generated by {\texttt{EllipSect}} for this galaxy.

\begin{figure}[!t]
\centering
  \includegraphics[width=\columnwidth,trim=10 10 10 10,clip]{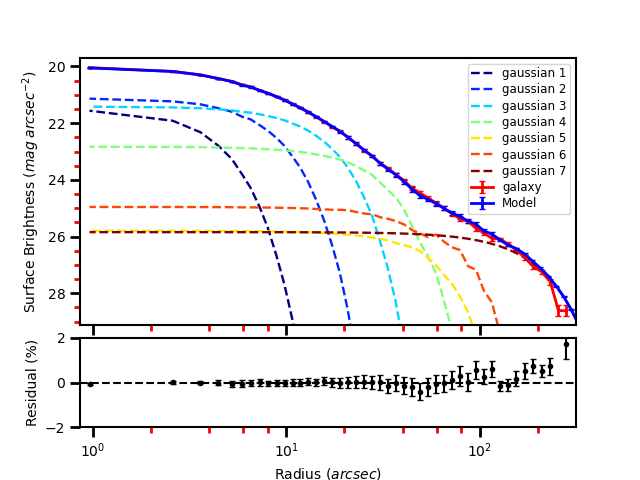}
  \caption{Example of {\texttt{EllipSect}} output for the cD galaxy Holm 15a and its model. 
  It shows a seven-Gaussian component fit. The red continuous line represents the galaxy and the blue 
  continuous line represents the multicomponent \texttt{GALFIT 3} model. Each Gaussian component 
  is shown as a dashed line and is color-coded, as indicated in the inset box at the top right. 
  The plot shows average SB vs. radius along the major axis. The residual percentage is displayed in the bottom plot. } 
\label{figa85}
\end{figure}

In Figure \ref{figa85}, we present the single-plot of the 
SB of Holm~15A. Seven Gaussians were used to cover the whole SB profile. The plot also shows the percent error. 
This figure was generated automatically by {\texttt{EllipSect}}, which computes the SB profiles of the galaxy, model, and individual components.

\subsection{2MASS: M61}

\begin{figure}[!t]
  \includegraphics[width=\columnwidth]{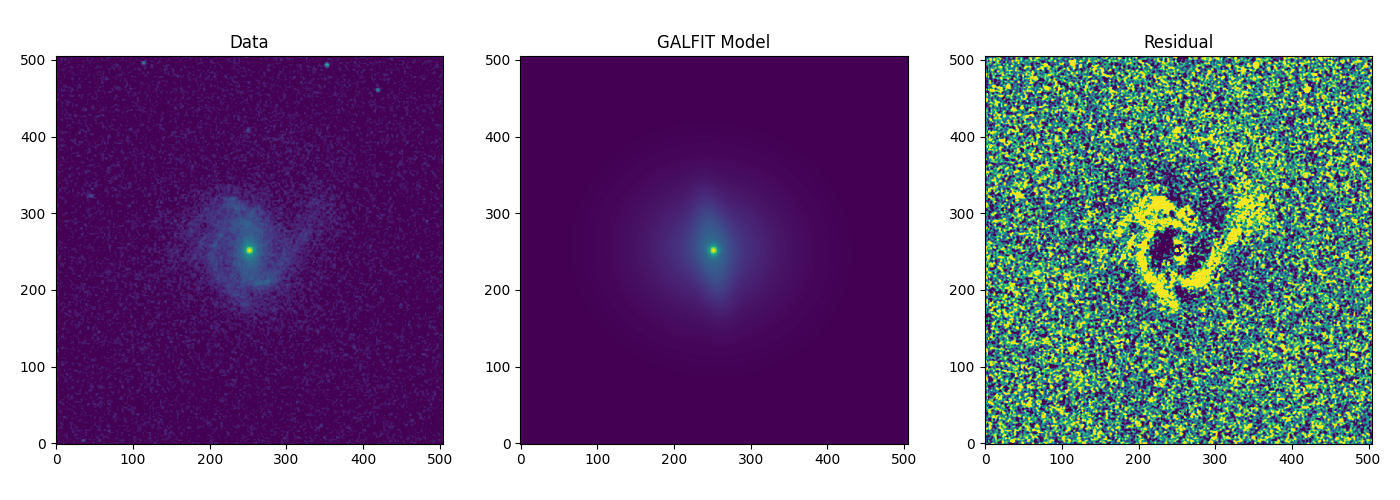}
  \caption{Example of  cube image generated by {\texttt{EllipSect}} for M~61
  galaxy. Details for this image are the same as the caption of Figure \ref{cube}.}
\label{cube2}
\end{figure}

In Figure \ref{cube2}, we show the galaxy, model, and residual for galaxy M~61 in the \textit{Ks} band of 2MASS.

\begin{figure}[!t]
  \includegraphics[width=\columnwidth]{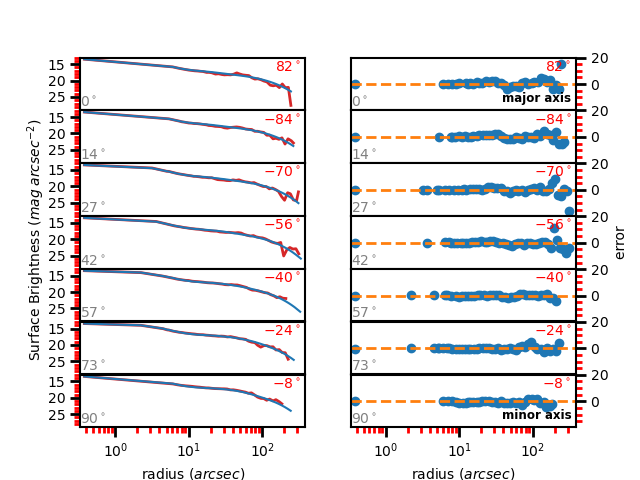}
  \caption{Example of {\texttt{EllipSect}} multi-plot output for M~61 galaxy and its model. 
This figure shows multiple plots of galaxy SB and model at different angles from the major axis 
(the major axis is the one with $0\deg$). The error is shown on the right hand side of the multiplot. 
  The angle in red at the top right indicates the angle measured relative to the image's Y-axis. 
  On the other hand, the subtle grey angle at the bottom left indicates the angle measured from the 
  major axis, starting from the top, with the major axis and the bottom plot with the minor axis.}
\label{figm61}
\end{figure}

Figure \ref{figm61} shows the multiplot of the fitting 
of M~61. This model consists of a series of S\'ersic and Exponential components. In
the region where the model fails to reproduce the SB model
of the galaxy, it results in apparent deviations.  

\subsection{HST: NGC4261}

\begin{figure}[!t]
  \includegraphics[width=\columnwidth]{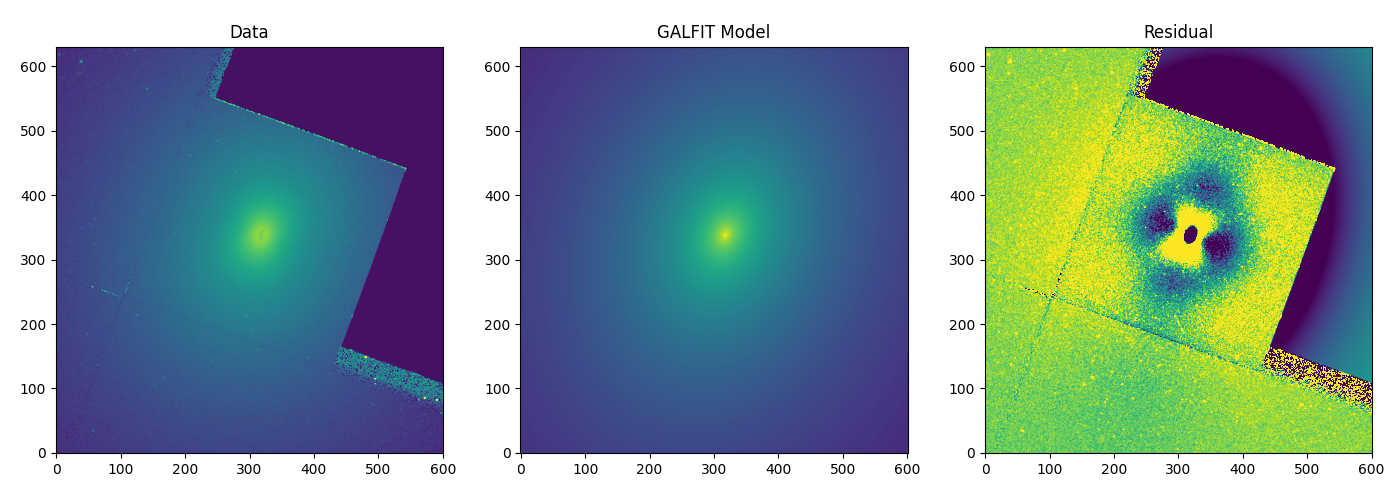}
  \caption{Example of  cube image generated by {\texttt{EllipSect}} for NGC4261
  galaxy. Details for this image are the same as the caption of Figure \ref{cube}.}
\label{cube3}
\end{figure}

In the last example, we display an HST image of NGC 4261. Figure \ref{cube3} includes the actual galaxy, its model, and its residual. 

\begin{figure}[!t]
  \includegraphics[width=\columnwidth]{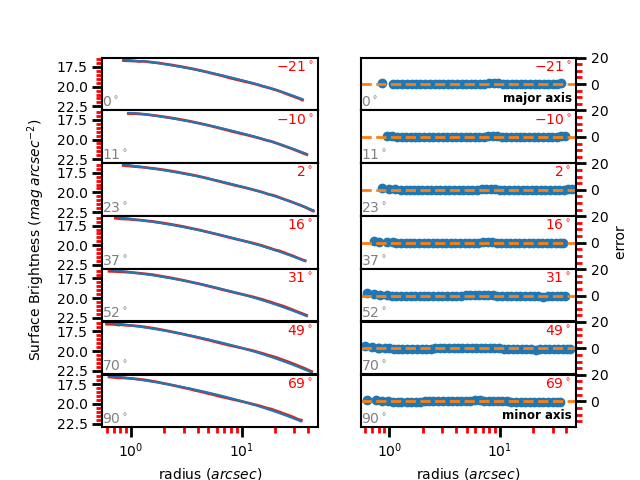}
  \caption{Example of {\texttt{EllipSect}} output for NGC4261 galaxy and its model. The 
  description of the image is the same as Figure \ref{figm61}.}
\label{figngc4261}
\end{figure}

Figure \ref{figngc4261} shows the galaxy NGC4261 taken with the HST using filter F547M. 
This model used a single Sérsic component to fit the galaxy's SB.

\section{Discussion}

\texttt{EllipSect} was developed to generate surface brightness profile plots, 
calculate non-parametric measurements, and export data derived from GALFIT model outputs.
Its primary purpose is to test the reliability of 
\texttt{GALFIT 3} 's model at various radii and angles for a given galaxy. In 
addition, {\texttt{EllipSect}} allows users to include individual components 
of the \texttt{GALFIT 3} models in the plots, enabling verification that 
these components behave as expected. This feature streamlines the analysis 
process by eliminating the need to switch between different program formats, 
making it practical to quickly confirm that the \texttt{GALFIT} model is 
appropriate for the galaxy.  \texttt{EllipSect} aids \texttt{GALFIT 3} users with a readily accessible and easy-to-use tool for generating accurate SB measurements and publication-quality plots.

One typically uses IRAF's \texttt{ellipse} in the conventional approach to generating SB profiles. This routine fits an ellipse to each isophote at every radius, from the central region to the galaxy's outskirts. As the radius changes, the ellipse's ellipticity and position angle may also vary.

One caveat of {\texttt{EllipSect}} is that it maintains the ellipticity and the angular position fixed throughout the entire galaxy, overlooking any possible variations of these parameters with radius. However, this 
fixed-parameter approach is advantageous when dealing with galaxies fitted simultaneously using 
\texttt{GALFIT 3}. In such cases, masking or removing the galaxy is insufficient, as the galaxies 
are blended on the line of sight, and it is challenging to distinguish where one galaxy ends, and the 
other begins \citep[e.g.,][]{haussler07}. In addition, {\texttt{EllipSect}} 
avoids the need for ``double'' fitting, i.e., two-dimensional 
and one-dimensional fitting, which IRAF's \texttt{ellipse} requires. 
One-dimensional routines may not be capable of simultaneous 
fitting isophotes for two neighboring galaxies, making simultaneous fitting difficult or impossible. While {\texttt{EllipSect}} is not a complete substitute for IRAF's \texttt{ellipse}, it offers alternative techniques to address the limitations of isophotal fitting.

Despite the widespread use of IRAF's \texttt{ellipse}, alternative techniques such as {\texttt{EllipSect}} offer unique advantages for SB analysis, which may benefit astronomers seeking to enhance their understanding of galactic structures.

The parameter  $r_e^t$  is the effective radius of the entire galaxy, which is an essential parameter derived by {\texttt{EllipSect}}; $r_e^t$ is critical to assess metallicity gradients, effective velocity dispersion ($\sigma_{eff}$)  and other relevant information \citep[e.g.,][]{2023MNRAS.526.2479G} or comparisons with numerical models. However, no single component's $r_e$ can accurately describe an entire galaxy, unless we deal with pure disk or bulge galaxies.

Determining which aperture to use for a galaxy's photometric variables can be somewhat arbitrary. Different photometric quantities require a suitable radius for proper evaluation, and standardizing this across various galaxies is difficult. In the case of \texttt{GALFIT 3}, the reduced chi-squared ($
\chi^2_\nu$) is computed using all pixels within the fitting region. However, changing the size of this region can alter $\chi^2_\nu$ and may include pixels from other objects, introducing additional noise. Consequently, the $\chi^2_\nu$ reflects the fit quality for every object within that region.

{\texttt{EllipSect}} uses the radius containing 90\% of the total light to compute various photometric variables, $r_90$ which is calculated in the same fahion as the effective radius.
Considering the region that contains 90\% of the total light minimizes the contamination from external objects. {\texttt{EllipSect}} calculates the Tidal parameter, Bumpiness, AIC, BIC, and a local $\chi^2_\nu$ inside $r_90$.

{\texttt{EllipSect}} provides corrections for galactic extinction and surface brightness dimming, but does not perform k-corrections at this time. Users can still apply morphological-based corrections to the galaxy, although this may be imprecise for cluster galaxies whose colors can differ from those of field 
galaxies. In such situations, manual correction is recommended. There are many ways to generate  \texttt{k}-corrections, we found the tool \url{http://kcor.sai.msu.ru/}, which uses redshift and the color index of a given galaxy as input \citep{chilingarian12}.

{\texttt{EllipSect}} provides a helpful useful tool for wrapping code to automate \texttt{GALFIT 3} fitting process. It can efficiently compute the necessary photometric variables required by these codes, thereby allowing for rapid analysis of model outputs for multiple galaxies, which would otherwise be time-consuming. In fact, {\texttt{EllipSect}} has been also designed to be used in conjunction with the  Driver for \texttt{GALFIT} on Cluster Galaxies (\texttt{DGCG}) code (A\~norve et al., in preparation), which automates \texttt{GALFIT 3} fitting for galaxies in dense environments.

\section{Conclusions and Future Work}

This paper introduces \texttt{EllipSect} algorithm implemented in \texttt{Python} 
designed to analyze outputs from \texttt{GALFIT 3}. It generates surface brightness profiles for galaxies, 
models, and individual components, offering single-plot and multiplot visualizations. Additionally, 
{\texttt{EllipSect}} computes various photometric parameters, including light fractions for each component, $r_e^t$, $\left<\mu\right>_e$, $\mu_e$, absolute magnitudes, the luminosity radius $90\%$, the effective radius of the galaxy, the Kron radius, and the Petrosian radius. It also calculates Tidal and Bumpiness parameters, signal-to-noise ratios, and local $\chi^2_{\nu}$ values to assess differences between models and observed galaxies.

{\texttt{EllipSect}} offers two methods for computing the sky background and enables users to distinguish among different galaxy models by computing  Bayesian information criteria. The code can be easily adapted to other \texttt{GALFIT 3} wrapping codes and will be integrated with \texttt{DGCG} in future work. This project aims to apply {\texttt{EllipSect}} to analyze data images containing hundreds of galaxies, such as galaxy clusters.

{\texttt{EllipSect}} is a robust tool for analyzing \texttt{GALFIT 3} output and extracting extensive photometric information for galaxies. Its flexibility and versatility make it an essential tool for studying galaxy properties and morphology. Future development will focus on expanding its 
capacity to handle larger datasets and integrate
it with other \texttt{GALFIT 3} wrapping codes.

\begin{appendices}

\section{\texttt{GALFIT 3} built-in functions or profiles}

Below, we describe the built-in functions or profiles implemented in the last distribution of \texttt{GALFIT 3}\footnote{See \texttt{GALFIT 3} user's manual, downloadable from \url{https://users.obs.carnegiescience.edu/peng/work/galfit/galfit.html}}. 
 We cite the original sources and main applications of each profile. \citet{2020A&A...634A.109B} has shown that the Nuker profile is the most general profile, which includes the S\'ersic profiles as a special case; hence, we have ordered the profiles from the more general to the particular cases. Our aim in including the profiles in this appendix is to homogenize the notation with that commonly used in textbooks.

\begin{enumerate}

\item \textbf{Nuker \citep{lauer95}:} This is a general profile that includes three power-law slopes, namely: $\gamma$ (inner), $\alpha$ (sharpness of the transition from inner to outer slopes), and $\beta$ (outer), they are free to vary and help to account for the large cores of very luminous early-type galaxies, including cD galaxies. This profile is expressed as follows: 
\begin{equation} \label{nuker}
    I(r)=I_b\, 2^{(\frac{\beta -\gamma}{\alpha})}\left(\frac{r_b}{r}\right)^{\gamma} \left[ 1 + \left(\frac{r}{r_b}\right)^{\alpha}\right]^{(\frac{\gamma-\beta}{\alpha})},
\end{equation}
where $r_b$ is the break radius, defined as the location at which the slope of the surface brightness 
profile changes, and $I_b$ is the SB at $r_b$.

The work of \citet{2020A&A...634A.109B} showed that the S\'ersic profile is a unique subgroup of the Nuker function; hence, the de Vaucouleurs profile, the exponential profile, and the Gaussian profile discussed below become specific cases of the Nuker profile, as well. 

 \item \textbf{S\'ersic \citep{sersic68}:} This profile was introduced as a generalization of the de Vaucouleurs profile and is typically used to fit bulges, bars, and 
 elliptical galaxies by allowing the concentration index $n$ to vary:

\begin{equation}
I(r) = I_e \exp\left[ -\kappa_n \left(\left( \frac{r}{r_e} \right)^{\frac{1}{n}}-1\right)\right],
\label{eq:sersic}
\end{equation}
where $I_e$ is the SB at the effective radius, $r_e$, which encloses half of the luminosity. The parameter $\kappa_n$ is related to $n$, so  $r_e$ becomes the half-light radius; for $n \in \mathbb{R}$, $\kappa_n=2n-\frac{1}{3}+\frac{4}{405n}+\frac{46}{25515n^2}+O(n^{-3})$ gives an approximation \citep{1999A&A...352..447C}. See, for example,  \texttt{GALFIT 3} user's manual and \citet{graham05} for a full discussion on the properties of the S\'ersic profile.

 \item \textbf{de Vaucouleurs \citep{devau48}:} A particular case of the S\'ersic 
  profile (index $n=4$, $\kappa_4= -7.66925$).
  A general tendency shows that  $n$ grows with galaxy luminosity; however, at high luminosities the tendency levels off ($ n\rightarrow 4$) 
  \citep[e.g.,][]{2003ApJ...594..186B,2011RMxAC..39..100L,2025arXiv250202546R}. Thus, as originally found 
    by \citet{devau48}, this profile historically provided a widely used approximation to 
the SB profiles of giant ellipticals. However, recent studies advocate using flexible S\'ersic indices (n varying) 
    for more accurate descriptions of elliptical galaxies' diverse structures \citep{graham05, kormendy09}. This 
profile has the following expression: 

\begin{equation}
    I(r) = I_e \exp\left[-7.66925 \left( 1- \left(\frac{r}{r_e} \right)^{1/4}\right)\right]
\end{equation}

\item \textbf{Exponential \citep[][]{1940BHarO.914....9P}:} This popular profile is used to fit the disks of spiral galaxies; it's a particular case of the S\'ersic profile with $n=1$:
\begin{equation}
  I(r) = I_0 \exp\left(-\frac{r}{r_s}\right),  
\end{equation}
 $I_0$ is the central SB and  $r_s$ is called the scale length, which is related to the effective radius by  $r_e=1.67835\,r_s$ for $n=1$. \citet{1970ApJ...160..811F} proposed a combination of a de Vaucouleurs bulge and an exponential disk to fit late-type galaxies. 
 
\item \textbf{Gaussian profile} is used to fit PSF profiles, the contribution of active galactic nuclei and bars; it's a special case of the S\'ersic profile when $n=0.5$. The following equation describes it:
\begin{equation}
 I(r)=I_0 \exp\left( \frac{-r^2}{2\sigma^2}\right),  
\end{equation}
The full width parametrizes the scale at half maximum ${\rm FWHM}= 2\sqrt{2\ln{2}}\approx 2.355\,\sigma$, instead of $r_e$.

\item \textbf{Edge-on disk \citep{1981A&A....95..105V}:} This profile fits the vertical structure of edge-on spiral galaxies. It results after considering an locally isothermal thin disk \citep{1981A&A....95..105V}:  

\begin{equation}
    I(r,z)=I_0 \left(\frac{r}{r_s}\right)K_1\left(\frac{r}{r_s}\right)\sech^2\left(\frac{z}{z_h}\right),
\end{equation}
where $I_0=I(0,0)$ is the central SB, $r_s$ is the scale length on the disk's major axis, $z_s$ is the scale height on the $z$ axis perpendicular to the disk, and $K_1$ is the modified Bessel function of the second kind.   

\item \textbf{Ferrers\footnote{\citet{peng10} called it modified Ferrer profile.}}: This model is named after the triaxial potential introduced by Norman Macleod Ferrers \citep[][pp. 61-62]{Ferrers1877, 1987gady.book.....B}. The modified Ferrer profile programmed in  \texttt{GALFIT 3} is given by the following expression: 
\begin{equation}
    I(r)=I_0\left[1-\left(\frac{r}{r_{out}}\right)^{(2-\beta)}\right]^{\alpha},
\end{equation}
which shows an almost flat core and an outer truncation, and whose  sharpness of the truncation is  set by $\alpha$, while the central slope is determined by $\beta$,
 $I_0$ is the central SB. This profile is only valid for $r\leq r_{out}$, i.e.,  $I(r\geq r_{out})\equiv0$.
 
This profile has been used to fit bars and lenses; it has eight free parameters. Notwithstanding, a S\'ersic profile whose index is $n<0.5$ approximates the Ferrers profile; therefore, either profile can be used to model the SB of bars. Besides, the results are indistinguishable \citep[e.g.,][]{2015ApJ...799...99K}.
 
\item \textbf{Empirical King \citep{1999glcl.conf..209E}:} It is given as follows: 
\begin{equation}
 I(r)=I_0 \left(1-\frac{1}{\psi^{\frac{1}{\alpha}}}\right)^{-\alpha}\left(\frac{1}{\xi^{\frac{1}{\alpha}}}- \frac{1}{\psi^{\frac{1}{\alpha}}}\right)^{\alpha};   
\end{equation}
where $\psi=\left( 1 - \frac{r_t}{r_c}\right)^2$ and $\xi=\left(1 - \frac{r}{r_c}\right)^2$, while the core radius $r_c$, the truncation radius $r_t$, and the power law index $\alpha$ are free parameters; however, in the standard modified King profile, $\alpha=2$.  This profile fits the SB of globular clusters.

\item \textbf{Moffat \citep{1969A&A.....3..455M}} is a profile used to fit PSF profiles or the emission of active nuclei of galaxies. The following equation describes it:
\begin{equation}
    I(r)=\frac{I_0}{\left( 1+ \left(\frac{r}{r_d}\right)^2\right)^n},
\end{equation}
where $n$ is the concentration index; as for the Gaussian profile, the scale is given by $FWHM=2\left(2^\frac{1}{n}-1\right)^\frac{1}{2}\,r_d$. Nevertheless, the Gaussian profile is more widely used than the Moffat. 

\item The user could provide the PSF as an image. However, care should be taken when selecting the stellar image's centroid, intensity, and sampling (see \texttt{GALFIT 3} user's manual).

\item \textbf{The background sky} is modeled as a plane, with an optional gradient applied along the $x$ and $y$ directions. 

\item \textbf{Spiral modes} are used to model the spiral arms or asymmetrical galaxies; see \citet{peng10} and  \texttt{GALFIT 3} user's manual for more details. 
\end{enumerate}

\section{How to install and run \texttt{EllipSect}}

\subsection{Installing \texttt{EllipSect}}

\noindent \texttt{EllipSect} is implemented in Python 3 for the analysis of GALFIT's outputs 
and the generation of scientific figures, and is available at \url{https://github.com/canorve/EllipSect}

\subsubsection{Quick installation}

\noindent The quickest way to install {\texttt{EllipSect}} is just by typing this into the 
command line:

\begin{lstlisting}
pip install ellipse
\end{lstlisting}

\subsubsection{Step-by-step installation}

The following libraries are required to install {\texttt{EllipSect}} (v3.4.4) along with \texttt{Python} $>=$ v3.8 \citep{python2009}:

\begin{itemize}
\item \texttt{numpy} $>=$ v1.20.3 \citep{Numpy2020}
\item \texttt{astropy} $>=$ v5.1 \citep{astropy:2013,astropy:2018}
\item \texttt{scipy} $>=$ v1.5.2 \citep{Scipy2020}
\item \texttt{matplotlib} $>=$ v3.5.2 \citep{matplotlib2007}
\item \texttt{mgefit} $>=$ v5.0.13 \citep{cappellari02}
\item \texttt{sphinx} $>=$ v3.2.1 \citep{sphinx2021}
\end{itemize}

After installing the libraries, download the latest release from the link shown above, and  install it via:

\begin{lstlisting}
cd EllipSect
pip install . 
\end{lstlisting}

or 
\begin{lstlisting}
cd EllipSect
python setup.py install
\end{lstlisting}

\noindent To check that everything is working properly, run the automated tests:

\begin{lstlisting}
python -m pytest 
\end{lstlisting}

\noindent Although \texttt{GALFIT} is not strictly required, it will require to create the model 
components and sigma image. Ensure you can call \texttt{GALFIT} from the command prompt. Otherwise, 
the automated tests will fail. \\

\subsection{Running \texttt{EllipSect}}
\label{a2}

\textbf{Easy run}: Once installed, run the command {\tt \texttt{EllipSect}} in 
the same directory that you run {\texttt{GALFIT}}. {The easiest way to run the program is:}

\begin{lstlisting}
EllipSect GALFIT.01
\end{lstlisting}

\textbf{Note}: {\texttt{EllipSect}} takes the \texttt{{\texttt{GALFIT}}} 
  output files ({\texttt{GALFIT.nn}}) where \texttt{nn} is the output file number you want to plot.

\textbf{More options}: As explained in the paper, \texttt{EllipSect} uses 
an ellipsed-based grid to compute the SB. The information
of this ellipse is obtained from the \texttt{GALFIT} output file, but if the user wants 
to provide their ellipse parameters, this is done via: 

\begin{lstlisting}
EllipSect GALFIT.01 -q 0.5 -pa 80
\end{lstlisting}

In this last example, an ellipse with a $0.5$ axis ratio and position 
angular of $80$ (measured from Y-axis) was used to create the plots. 

In case your model contains two or more components, the
\texttt{comp} (or \texttt{cp}) option will plot each component (for an example of this check figure \ref{figa85}):

\begin{lstlisting}
EllipSect GALFIT.01 -cp
\end{lstlisting}

There are many options to change the format 
of the plots such as: {\tt logx, ranx, rany, pix, grid, dpi, galax}, 
but if the user prefers to make their plots the option {\tt sbout}
will create a file (or files) containing the points of the 
the SB of the galaxy and model at each radius.

To compute the photometry of the galaxy and model such  as  
{\tt snr, BIC, AIC, absolute Magnitude, Luminosity, Bumpiness, Tidal
} and other photometric variables the option {\tt phot} creates 
a file with the computed variables:

\begin{lstlisting}
  EllipSect GALFIT.01 --phot 
\end{lstlisting}

{\texttt{EllipSect}} has the option to compute the sky level to compare with or to provide to 
{\texttt{GALFIT}}. As explained above, the program uses two methods to compute it: The 
gradient method and the random box method. To compute the gradient method: 

\begin{lstlisting}
  EllipSect GALFIT.01 -gsky
\end{lstlisting}

And to compute using the random box method: 

\begin{lstlisting}
  EllipSect GALFIT.01 -rsky
\end{lstlisting}

The ring's width and box's size can be modified 
from their default options.

For more options and to see how to use them, type:

\begin{lstlisting}
EllipSect --help 
\end{lstlisting}

\subsubsection{Running \texttt{EllipSect} as a script}
If you want to use {\texttt{EllipSect}} inside your \texttt{Python} script, you can call it in the following way:

\begin{lstlisting}[caption=Example Python Code with \texttt{EllipSect}]
    from ellipsect.inout.read import ArgParsing
    from ellipsect.sectors.sect import SectorsGalfit

    args=['GALFIT.01','--logx', '--phot','--noplot']
    parser_args = ArgParsing(args)
    photapi = SectorsGalfit(parser_args)

    print("Akaike Criterion: ", photapi.AICrit)
    print("Bulge to Total: ", photapi.BulgeToTotal)
\end{lstlisting}

A manual and more details can be found in \url{https://github.com/canorve/EllipSect}

\end{appendices}

\bibliography{ref}

\end{document}